# Gestion efficace de séries temporelles en P2P:

## Application à l'analyse technique et l'étude des objets mobiles

GEORGES GARDARIN, BENJAMIN NGUYEN, LAURENT YEH,
KARINE ZEITOUNI, BOGDAN BUTNARU, IULIAN SANDU-POPA

Laboratoire PRiSM, 45 avenue des Etats-Unis, 78035 Versailles Cedex
*firstname.lastname@prism.uvsq.fr*

**Résumé.** Dans cet article, nous proposons un modèle simple et générique pour la gestion des séries temporelles. Une série temporelle est composée d'un calendrier avec une valeur typée attribuée à chacune des dates. Bien que le modèle supporte n'importe quelle sorte de valeur XML typée, nous nous intéressons dans cet article aux nombres réels, qui sont le cas d'applications pratiques rencontrées le plus souvent. Nous définissons des opérations classiques d'espace vectoriel (plus, moins, multiplication scalaire) et également des opérateurs de type relationnel ou métier pour la gestion des séries temporelles. Nous montrons l'intérêt de ce modèle générique sur deux applications : (i) un système d'analyse technique des cours de bourse, (ii) un système d'évaluation de conduite écologique. Ces deux applications présentent des particularités différentes : l'analyse technique demande des opérations de fenêtrage poussées, tandis que la gestion des transports requière des requêtes complexes. Notre modèle a été implémenté et testé en PHP, Java et XQuery. Nous donnons dans cet article des résultats illustrant des temps de calcul de plus de 5000 séries de plus de 100.000 valeurs, ce qui est un cas d'utilisation courant pour ce genre d'applications. Les résultats montrent que ce calcul est laborieux sur un seul PC unique, même très puissant. Ainsi, dans le cas d'une communauté d'utilisateurs qui partagent les calculs sur les séries temporelles, nous introduisons une implémentation P2P Java de ces séries temporelles en les divisant en segments, et nous proposons des algorithmes optimisés pour le calcul de certains opérateurs.

**Abstract.** In this paper, we propose a simple generic model to manage time series. A time series is composed of a calendar with a typed value for each calendar entry. Although the model could support any kind of XML typed values, in this paper we focus on real numbers, which are the usual application. We define basic vector space operations (plus, minus, scale), and also relational-like and application oriented operators to manage time series. We show the interest of this generic model on two applications: (i) a stock investment helper; (ii) an ecological transport management system. Stock investment requires window-based operations while trip management requires complex queries. The model has been implemented and tested in PHP, Java, and XQuery. We show benchmark results illustrating that the computing of 5000 series of over 100.000 entries in length – common requirements for both applications – is difficult on classical centralized PCs. In order to serve a community of users sharing time series, we propose a P2P implementation of time series by dividing them in segments and providing optimized algorithms for operator expression computation.

# 1. Introduction

Research in time series[1] (noted TS for short) [12][14] has been very prolific in the last decade. Several domains of applications such as medicine, finance, economy, climate evolution, transport control have emerged. Time series can be used to model temporal, or even spatio-temporal data. Several tasks have been considered including query by content, pattern discovery, trend computation, summarization, dimensionality reduction, classification, segmentation, etc.

Many authors have proposed distance computation techniques for mining series and indexing time series based on distances. As a consequence, numerous summarization methods for dimensionality reduction and efficient distance measures have been introduced. Thus, dimensionality reduction has focused attention applying techniques such as Discrete Fourier Transform, Discrete Wavelet Transform, Singular Value Decomposition, Piecewise Linear Approximation, etc. Common benchmarks are required to evaluate these methods. See for example [7] for a comparative analysis of representations and distance measures.

In most approaches, a time series is a sequence of (time, value) pairs. The problem when comparing time series and computing distances between them is that the sequence can be very long and somehow biased by missing values, noise, or other phenomena. From a practical point of view, it is unrealistic that mapping time series to smaller spaces will be sufficient for detecting and correcting these biases. We claim that performing basic operations and transformations on time series in parallel is another approach that has not been fully explored yet. This might be due to the fact that no common model including standard operations (somehow similar to relational algebra) has emerged for processing time series. Several DBMSs include them as an SQL3 data type, but with different and non standard functions.

Our contributions in this paper are the following: (i) we propose a basic model for time series as a sequence of (time, value) pairs with an extensible set of operations and a finite (but extensible) calendar. The basic set of operations are vector space operations (+, -, scalar multiplication) plus relational operations (filter, join, union, intersection) adapted to time series. Time is discrete and known time values are shared between various series through a common calendar giving all considered values in time (which is itself a time series). The model is extensible in the sense that application oriented operators can be defined and included in the model. (ii) We illustrate the use of this model on stock market quotes where window sliding operations are common, and logs of vehicle on-board sensors where complex integral-based operations are frequent. (iii) We have experimented several main memory implementations on application data from both examples. (iv) Based on these experiments, we show that an efficient way for processing long time series (e.g., with more than 100.000 entries) is to divide them and distribute them in a P2P network. We describe our P2P implementation with a generic optimizer currently in progress, and show some first performance measures. The system can parallelize both the basic model and the application-oriented operators.

The uniqueness and novelty of our contribution resides in the fact that (i) we provide an extensible model with functional operators easily incorporable in XQuery 1.1 or to be used in a specific implementation. (ii) We propose a P2P implementation with comparative experimentation of our prototype on real application data. To our knowledge, no P2P implementation of a time series engine has been developed so far.

---

[1] Time series considered in this article should not be confused with data streams [3][17][18][11], although some similarities and common applications exist (see Section 6). The main difference is that a Time series is *persistent* data queried by *on demand* client requests while data streams are *transient* data queried by *continuous* queries. The former are OLAP oriented while the latter are real time oriented (event detection). Time series manage historical data while data streams manage data fluxes. There is a certain overlap at the frontier between time series and data streams that we do not consider here.

The rest of this paper is organized as follows. Section 2 reviews the concept of time series, and gives an overview of our basic model, including calendar, null values, vector space and relational operators. Section 3 introduces two possible applications of this work: stock market investing and ecological transport analysis. Section 4 briefly describes three centralized implementations in PHP, Java, and XQuery. We compare them with a simple application-driven benchmark. Section 5 describes the P2P time series engine and an experimentation of its performance and gains over the centralized systems of the previous section. In Section 6, we review recent work geared specifically towards time series models and P2P processing. Section 7 concludes the paper and discusses possible future extensions of our work.

## 2. Time Series Model

Our basic model is derived from the Roses project [23] adapted and extended for our needs. The model is composed of a vector space of time series equipped with relational-like operations mapping one, two or more time series to one. The model also includes aggregate operators to change the time unit of a series that we do not detail due lack of space. The model also encompasses window-based operations similar to those proposed in the current working draft of XQuery 1.1 [28].

### 2.1 Vectors and Vector Space

We define a **time series** as a potentially infinite vector of values. In the rest of the article, we use $n$ to denote the length of the time series. The vector is associated with a **calendar** giving for each point in time[2] the index of the entry. Time can be of different granularities (e.g., second, day, hour, and week). While in general any kind of XML type, in this article, due to application requirements, values are **double precision floats**. The calendar starts at a given time which corresponds to the first entry in the associated series; all time units from start to end (the last recorded entry) correspond to an entry. An item is a couple (time, value), i.e., a row in the vector. There exists two possible and distinct null values, the empty (or non-exist) value (denoted "**!**") meaning that there is no value for the given time and the unknown value (denoted "**?**"); the first appears for instance after filtering a series while the second one may appear when the series is not totally computed yet or when computation leads to a division by 0.

Time series constitute a *linear vector space*, i.e., a mathematical structure formed by a collection of vectors that may be added (addition is denoted +) together and multiplied (multiplication is denoted *) by numbers, called scalars in this context. Scalars are real numbers in our case. Multiplication and addition of null values are defined as follows, s being a real:

$$\text{(i)} \begin{cases} !+!=! \\ !+?=? \\ ?+?=? \end{cases} \qquad\qquad \text{(ii)} \begin{cases} s*!=! \\ s*?=? \end{cases}$$

Time series can be combined linearly in expressions such as $TS_1 + s * TS_2$ where $TS_1$ and $TS_2$ are time series (in practice of same calendar and dimensions). They have all properties of vector spaces: addition is associative and commutative, has an identity element and existence of null and identity vectors. Scalar multiplication is distributive with respect to vector addition, etc. These properties are interesting for query optimization.

The time series linear vector space is also a *metric space*. Thus, in the last decade, there has been an increasing amount of interest in time series representation methods and distance measures. Many dimension reduction methods and over a dozen distance measures have been proposed. Most of them are nicely compared in [7]. The most popular are Euclidian distance [10] and Dynamic Time Warping [25]. Representation methods preserving distance ranking are very important for similarity search and mining tasks such as classification or clustering. In this paper, we do not focus on distance computation, although we have implemented some. We plan to integrate in our query processing engine vector dimensionality reduction operations and complex distance functions in the near future for other applications.

---

[2] There are various possibilities to implement time. Our P2P Java implementation is based on ISO 8601, with *arbitrary* precision. XML for instance demands at least *ms* precision. Due to lack of space we can not detail further.

## 2.2 Relational Operators

Logical operators are derived from relational algebra operators specialized for time series. First, the model includes the counterpart of the selection and projection relational operations (also called filter and map). The result of a selection applied to a time series is a time series, keeping the original value if the predicate is satisfied and replacing it by the empty value (!) if not. More formally, denoting [t,v] the entry t of value v of the processed time series :

$$\text{SEL}_{\text{pred}}(S) = \{[t, v] \mid [t, val] \in S \wedge v = pred(val)\}$$

where $pred$(val) = val if val satisfies the predicate *pred* and ! otherwise.

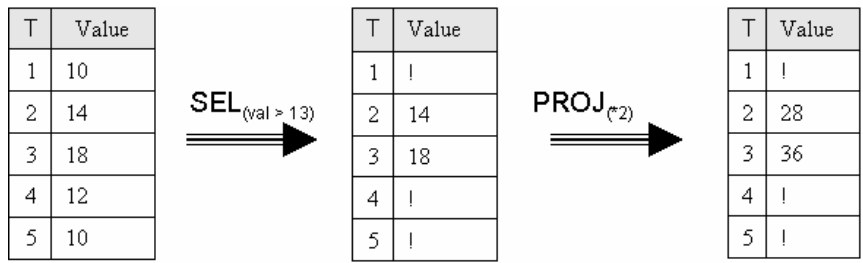

Fig. 1 - Examples of Selection and Projection

A projection applies a similar transformation defined by a function to each value of the time series it is applied on. More formally:

$$\text{PROJ}_{\text{fun}}(S) = \{[t, m] \mid [t, val] \in S \wedge m = fun(val)\} \qquad (1)$$

Examples of Selection and Projection of time series are given in Fig. 1. Note that operators can be composed as with relational algebra to form algebraic expressions. This is true for all operations of our time series algebra. Furthermore, the map function must be defined on null values (giving null simply by default).

The model also includes some adaptation of the relational outer union and intersection, simply called **union** and **intersection**. They are formally defined as follows, and illustrated in Fig. 2:

$$S_1 \cup S_2 = \{[t, v] \mid [t, v] \in S_1 \vee [t, v] \in S_2\} \qquad (2)$$

$$S_1 \cap S_2 = \{[t, v] \mid [t, v] \in S_1 \wedge [t, v] \in S_2\} \qquad (3)$$

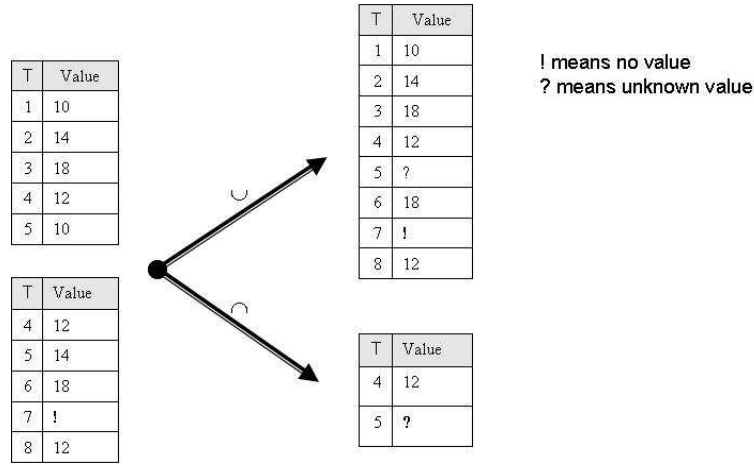

Fig. 2 - Example of Union and Intersection

Finally, we introduce a k-ary **join** operation for time series based on the same calendar. This operation performs a join on the time attributes of $k$ time series using the same calendar, and then applies a mapping function to the tuple of values of the $k$ time series. More formally:

$$\text{JOIN}_{\text{fun}}(S_1, \dots S_k) = \{[t, m] \mid [t, val_1] \in S_1 \wedge \dots [t, val_k] \in S_k \wedge m = \text{fun}(val_1, \dots val_k)\}. \quad (4)$$

This operation is very useful for applications computing derived data from several time series. Notice that it can be computed in linear time since the time series have the same calendar, thus entry $i$ corresponds to time $i$ everywhere[3]. Joins on values are also possible, but so far we have not see their uses in the considered applications.

# 3. Application Oriented Extensions

We provide a time series server for processing time series imported from relational databases or XML format. The time series server implements the basic extensible model as described above, which includes operations useful for all applications. In this section we discuss application specific extensions.

### 3.1 Stock Selection and Strategy Evaluation

Stock investing is a quite difficult task, as shown by the recent crisis. Time series can be very useful for implementing and testing investment strategies on past data. Stock investing covers a broad range of stock analysis techniques. Experts distinguish fundamental and technical analysis. Fundamental analysis involves the study of company business and earnings. Technical analysis attempts to consider stock prices and volumes as temporal signals and to analyze these signals based on indicators, patterns, or events. We selected technical analysis as a first application for our time series model.

Most indicators used in technical analysis are window-based, i.e., compute a summary of a sliding window of values. That means that the $i$th indicator value is a function of the $w$ previous ones, $w$ being the window size. In the rest of the article, we use $w$ to denote the length of a window. Thus, guided by our applications requirements, we enrich the space of time series with a generic window-based operation:

---

[3] This is the main reason behind explicitly indicating non-existent values using the **!** symbol. Of course, non-existent values could simply have their date removed altogether, but with a slight increase in computational time. These considerations are somewhat out of the scope of the paper.

$$\text{WIN}_{\text{fun}}(S) = \{[t, val] \mid val = fun([t-1, val_1], [t-2, val_2], \dots [t-w, val_w])\} \qquad (5)$$

Let us recall that $[t-i, val_i]$ designates the entry $t-i$ of the time series S of value $val_i$ ; if $t-i$ is negative, $val_i$ is set to $val_0$.

Popular window-based operations in stock analysis are the Moving Average (MAVG) and the Relative Strength Index (RSI). The MAVG computes the classical moving average series of a series S with a sliding window of size w. Let $V = \text{MAVG}_w(S)$. The value $V[t]$ of entry t is defined by the equation[4]:

$$V[t] = \sum_{t-w}^{t} \frac{S[t]}{w} = \frac{(w-1) * V[t-1] - S[t-w] + s[t]}{w} \qquad (6)$$

A variation is the exponential moving average[5] where value $[t-i]$ is moderated by a weight $(1-alpha)^i$. The RSI is a popular technical momentum indicator that compares the magnitude of recent gains to recent losses in an attempt to determine overbought and oversold conditions of an asset. $R = \text{RSI}(S)$ is calculated using the following formula: $R[i] = 100 * \dfrac{G[i]}{G[i] + L[i]}$ where $G[i]$ is the sum of the gains in the window from $S[i-w]$ to $S[i]$ and $L[i]$ is the absolute sum of the losses in the same window.

There exist several similar indicators, as for example the Moving Standard Deviation, the Moving Maximum, the Moving Linear Regression Gradient, the Momentum (MOM), the ADX, and many others [15]. Let us stress that the considered operators generate a **result** time series from the initial time series based on the same calendar; giving for each time *t* the relevant indicator. This is useful for developing strategies as introduced below.

We note $C_{op}^{w}$ the cost of computing operator *op* on a window of size *w*. $C_{op}^{w}$ is usually polynomial in *w*. Most common complexities are O(1) and O(*w*). We note $C_{op}^{TS}$ the cost of computing the whole time series. In general, $C_{op}^{TS} \propto C_{op}^{w} \times n$. Computation time experimental results are given in Section 4.

Other operators can be computed by combining logical, vectorial, and windowing operators. For example, a common indicator in stocks investment is the Moving Average Convergence/Divergence (MACD). It is one of the simplest indicators used by some investors. A usual formula for the MACD[6] is the difference between a stocks 26-day and 12-day moving averages. Usually, a 9-day moving average of MACD is computed to act as a signal line to buy or sell when crossing 0. The following expression computes the MACD of a series S, then the signal line from the MACD, and finally gives a non empty value supporting a buy decision:

$$\text{BUY} = \text{SEL}_{>0}(\text{MAVG}_9(\text{MAVG}_{12}(S) - \text{MAVG}_{26}(S))) \qquad (7)$$

If you are interested in stocks investment, it is recommended to use a (good) strategy, i.e., a condition for buying and a condition for selling. In general, strategies are based on both fundamental analysis and technical analysis, and often more. In this application, we are interested in *evaluating many different strategies* applied to *thousands* of different stocks over a *minute-by-minute sampling of ten years of data*, based on technical analysis only. Most technical strategies are expressed as events over indicators, e.g., the moving average 9 crosses over the MACD 12/26. With time series, strategies can be simulated and evaluated on the past for given time intervals (i.e., periods). A buy strategy leads to a time series giving the buy events while a sell strategy leads to a time series giving the sell events. In general, a strategy can be expressed as two queries (a buy and a sell query) on the time series associated with the considered stock. The queries include the quotes of a time interval (e.g., several years) which is a subset of the calendar. The calendar may be day-based (each open day is an entry) or minute-based (each open minute is an entry), depending on the type of analysis. A quote can comprise several time series (for instance a daily quote might have open, close, low, high, and volume values). For example, a sell strategy for the MACD could be to sell when the $MAVG_{26}$ becomes higher than the $MAVG_{12}$ by a factor of 10%:

$$SELL = SEL_{>1.1}(MAVG_{26}(S) / MAVG_{12}(S)))$$  (8)

---

[4] As you can see with the equation a *naïve* implementation would be O(*w*) but it is of course possible to compute $V[t]$ in O(1) time.
[5] Exponential moving average can also be computed fast using bit-shifting. However our focus in this paper is not the specific optimization of operators but global optimization.
[6] Note that depending on the sampling granularity, a 9 day moving average could be an average over 9 values for day granularity, 72 values for hour granularity (markets open 8:30 hours per day) or 4590 values for minute granularity.

In summary, the application requires in addition to the basic model window-based operators and the ability to run efficient complex expressions with stats operators on long series: a (non-leap) year of quotes at minute resolution makes series 183.600 entries. Queries are functional expressions to compute during a time interval, for example the French stock exchange from 2000/01/01 to 2009/01/01, i.e. 1.653.930 entries.

### 3.2 Ecological Driving Evaluation

Many studies and surveys use sensors to collect large-scale data, which results in huge time series databases. One application that we have studied [24], called *naturalistic driving survey* is related to ecological driving in transportation research. It consists of collecting a large amount of data (e.g., location, speed, acceleration and choice of gears) from a large number of drivers over an extended period of time in a natural situation. Then, various indicators are computed in order to describe drivers' behavior. Based on these indicators, exploratory data analysis, such as factorial analysis, is performed to evaluate their correlation or their influence on gas emissions. This may help, for instance, in establishing eco-driving instructions or more accurate traffic emission models and predictions.

Driving behavior can be described using many indicators; average speed being the most common. Ten standardized driving parameters are used to test vehicle emissions and fuel consumption in automotive laboratories, and many others (sixty two) have been used in [9]. Most of them require advanced filtering, and computations on time series. For example, average driving speed (excluding stops), or Relative Positive Acceleration (*RPA*) that represents the power demand, and combines speed and positive acceleration. RPA is defined as:

$$RPA = \frac{1}{x} \int v a^+ dt$$  (9)

where $v$ denotes the vehicle speed, $a$ its acceleration at a given time $t$ ($a^+$ restricts it to positive acceleration), and $x$ the total duration of the trip.
Another important parameter is Positive Kinetic Energy (*PKE*), which reflects the oscillations of speed curves. It increases with many local min and max of the speed curve. It is computed as:

$$PKE = \frac{\sum_{dv/dt>0} v_f^2 - v_s^2}{x} \text{ with } v_f = \text{final speed, } v_s = \text{start speed}$$  (10)

Many other indicators are of interest, among which: the Proportion of Standstill Time (*PST*), i.e. when the speed is < 2 km/h; mean duration of stops; the number of brakes per km; mean number of gear-shift per km; average engine speed (of RPM); average braking distance before a traffic light or a road curve; etc. Previous studies have found that *RPA*, *PKE*, and *PST* are the main impacting factors of driving behaviour that impact gas emissions.

Let us stress that in the context of Ecological driving, many of the results produced are scalar functions, i.e. functions that take as an input a time series and produce a scalar as result. Again, the complexity of the calculation of such functions is usually polynomial with regards to the length of the time series.

In this paper, we have implemented part of these indicators in order to assess the impact of an Intelligent Speed Adaptation (ISA) system on driving behavior and fuel saving. The ISA device is called LAVIA (from the French acronym "speed limiter that adapts to the legal speed limit"). Our tested dataset is related to a complementary study in the LAVIA project [8], and consists of eight trips of two drivers. Each person did the same trip four times. The duration of a trip is approximately 45mins and its length is approximately 47km. Environmental conditions (traffic congestion, weather, vehicle weight, engine temperature, etc.) were similar for each trip. For each driver, the four trips correspond to four driving styles: normal, nervous, economical and LAVIA. For a normal driving style, there are no specific recommendations. This style corresponds to a week-end leisure drive, when the driver is not in a hurry. For a nervous style, the driver should drive like when he is rushing and consider time factor as a priority, of course without taking any risk. The economical style intends to minimize the fuel consumption by pursuing the following recommendations: change to a superior gear as soon as possible (e.g. at 2500 rpm with a petrol engine); maintain a constant speed at the highest possible gear; anticipate the decelerations in order to avoid strong breaking; decelerate softly by releasing the gas pedal; cut the engine for all stops longer than a minute. Finally, LAVIA style corresponds to normal style with the LAVIA system active.

The time series studied in this paper contain the following information: road identifier, position (GPS provided latitude and longitude), legal speed, vehicle speed, acceleration and fuel consumption. The size of *each* time series is about 4000 measures; therefore for a given trip we had about 24.000 elements. It is important to note that the whole survey will contain many more attributes and thousands of different trips. Efficiency and scalability of the data management system are therefore crucial. We show in Section 4 some brief results concerning these calculations.

## 4. Centralized Implementations and Evaluations

In this section, we describe our centralized implementations of the model that we have done for evaluating the feasibility and comparing performance, and we also compare results with the Qizx XQuery engine, that we have also experimented on. We have chosen to use both our PHP and Java systems since all the operators have not yet been implemented in Java.

Time series are converted to the following simple XML format using XSLT, then loaded into the server.

<document>(<timeseries><date/><value/></timeseries>)*</document>

The date element is of type xs:date (ISO 8601) and the value element is an xs:anySimpleType., values used here were all xs:double. In practice, ! means no entry for that given point in time and ? is encoded as NaN. In principle, it would also have been possible to encode them as -NaN and +NaN.

We produced our data using various financial site [29][30][31], for our stock application and the Eco Transport relational database for our driving application.

### 4.1 Centralized testing environment

All centralized tests have been executed on a Xeon-X5450@3.00GHz with 4GB RAM (process RAM settings are indicated below) running Vista-64, using only a single processor core. We have also run tests using multiple cores that are beyond the scope of this paper (approximate gain when using 4 cores of a factor 2.5). Our Java implementation (and also Qizx) is running on version 1.6.0_14 (32 bit) with 1GB heap space (physical maximum when running 32 bit Java on Windows is 1.5GB).

### 4.1 Implementing the Model in PHP 3.0

Historically, our first implementation was done in PHP 3.0. A time series is a vector giving for each entry its value. The time series is described by meta-data including its functional name (the data source for a base time series or the tree of operations with parameters computing the series for a derived one) and the calendar. We could have separated the functional expression from the name, but we found convenient to integrate both in order to produce a unique DHT key (see Section 5). The calendar is a vector of time instants. Calendars are shared by time series with common time scale. Other meta-data is the length, the time unit, etc. Vectors are simply one-dimensional array of floats. Time series are cached by the PHP engine. The size of the cache is a PHP parameter (set to 1GB for benchmarks). The implementation of window operations was particularly simplified using dynamic arrays. The resulting library is simple and efficient as shown by the benchmark results.

### 4.2 Implementing the Model in Java

Our newer implementation is done in Java, in order to plug it into our P2P framework called [5]. This implementation is very generic and can be enriched simply by programming new classes that compute aggregate functions over a given window of a time series. The cost of using a generic model reduces its efficiency, therefore we have also implemented specifically optimized functions, that may need more information than just the window values. We show for instance the difference in time between computing using our generic function approach vs. specific implementation for WAVG.

### 4.3 Implementing the model in XQuery 1.1

W3C-XQuery 1.1, which is currently in public working draft status [28] introduces advanced support for time-series based on windows, based on the work of [4] who first proposed a generic construct (forseq) to query window sequences and continuous streams. Performance should be comparable to the performance of SQL with sequence data type operations: There should not be a performance penalty for using XQuery, as stated in [4]. Note that Qizx runs in multithreaded mode, therefore test times are less slightly more dispersed.

Concerning our data, we use directly time series converted to the schema show in Section 4. The functions that we have written in XQuery use this schema as a basis for the time series algebra and implement some of the operators of Section 2. One main difficulty is that math functions (such as exp) are absent from basic XQuery, and need therefore to be implemented as external functions, which is somewhat sub-optimal.

Our centralized tests have been run using Qizx 3.0 one of the only XQuery processors that already supports some 1.1 features. We are currently working on our P2P XQuery 1.1 processor (see [27] for information on it advancement) that will directly implement many missing functionalities for math oriented computing. To show how code is written in XQuery, we give as an example the MAVG operator computed in naïve linear complexity:

```
declare function ts:mavg($ts as ts:document, $i as xs:integer) as
ts:document{
<ts:document>{
for sliding window $w in $ts//ts:value
start at $s when fn:true()
only end at $e when $e - $s eq $i  -1
return
<ts:timeseries>
<ts:date>{(data($wn/preceding-sibling::date))[$i]}</ts:date>
<ts:value>{avg(data($wn))}</ts:value>
</ts:timeseries>
}</ts:document>
};
```

## 4.4 The benchmark

To evaluate the performance of the three implementations and determine the limits of these clever but straightforward main-memory implementations, we defined a simple benchmark composed of four queries, three dealing with stock investment and one with transport. We used the CAC40 quotations (called PX1) from 1990 to 2009 as data source, constructing time series of various lengths. We experiment with different window sizes, which are current in stock indicators: 10, 50, and 100. The queries are as follows:

(Q1) computes the moving averages, i.e., MAVG(PX1, w), where w is the window size.

(Q2) computes the RSI, i.e., RSI(PX1, w).

(Q3) computes the MACD, which is a difference between a short and a long exponential moving averages, more precisely MINUS(XAVG(PX1,3), XAVG(PX1, w));

(Q4) computes the positive kinetics energy of the CAC; this is indeed an eco-transport indicator that we apply to the CAC seen as a series of speeds. It is quite complex as given by the formula MINUS(MULT(MULT(PX1, SEL(MOM(PX1,2), >0)), MULT(PX1, SEL(MOM(PX1,2), >0))), MULT(MULT(SHIFT(PX1), SEL(MOM(PX1,2), >0)), MULT(SHIFT(PX1), SEL(MOM(PX1,2), >0)))).

| N | W | PHP | JAVA | QIZX |
|---|---|-----|------|------|
| 1000 | 10 | 4 | <1 | 16 |
| 1000 | 50 | 11 | <1 | 45 |
| 1000 | 100 | 19 | <1 | 91 |
| 2000 | 10 | 7 | <1 | 28 |
| 2000 | 50 | 22 | <1 | 90 |
| 2000 | 100 | 38 | <1 | 178 |
| 4000 | 10 | 15 | <1 | 53 |
| 4000 | 50 | 44 | <1 | 179 |
| 4000 | 100 | 79 | <1 | 357 |
| 16000 | 10 | 60 | 4 | 212 |
| 16000 | 50 | 176 | 4 | 765 |
| 16000 | 100 | 316 | 4 | 1404 |
| 100000 | 10 | 375 | 25 | 1914 |
| 100000 | 50 | 1097 | 25 | 5026 |
| 100000 | 100 | 1974 | 25 | 9251 |
| 500000 | 10 | 1875 | 128 | 9862 |
| 500000 | 50 | 5510 | 130 | 28259 |
| 500000 | 100 | 9888 | 129 | 49347 |

Tab. 1 WAVG computation time (in ms)

| N | W | PHP | JAVA | QIZX |
|---|---|-----|------|------|
| 1000 | 10 | 10 | 17 | 402 |
| 1000 | 50 | 29 | 31 | 401 |
| 1000 | 100 | 53 | 56 | 405 |
| 2000 | 10 | 22 | 15 | 1559 |
| 2000 | 50 | 62 | 60 | 1561 |
| 2000 | 100 | 114 | 114 | 1558 |
| 4000 | 10 | 45 | 43 | 6282 |
| 4000 | 50 | 127 | 123 | 6259 |
| 4000 | 100 | 230 | 241 | 6334 |
| 16000 | 10 | 198 | 141 | 97807 |
| 16000 | 50 | 526 | 490 | 99864 |
| 16000 | 100 | 934 | 917 | 96759 |
| 100000 | 10 | 1238 | 902 | out of memory |
| 100000 | 50 | 3291 | 3074 | out of memory |
| 100000 | 100 | 5838 | 5516 | out of memory |
| 500000 | 10 | 6196 | 4251 | out of memory |
| 500000 | 50 | 16457 | 14501 | out of memory |
| 500000 | 100 | 29209 | 27393 | out of memory |

Tab. 2 RSI computation time (in ms)

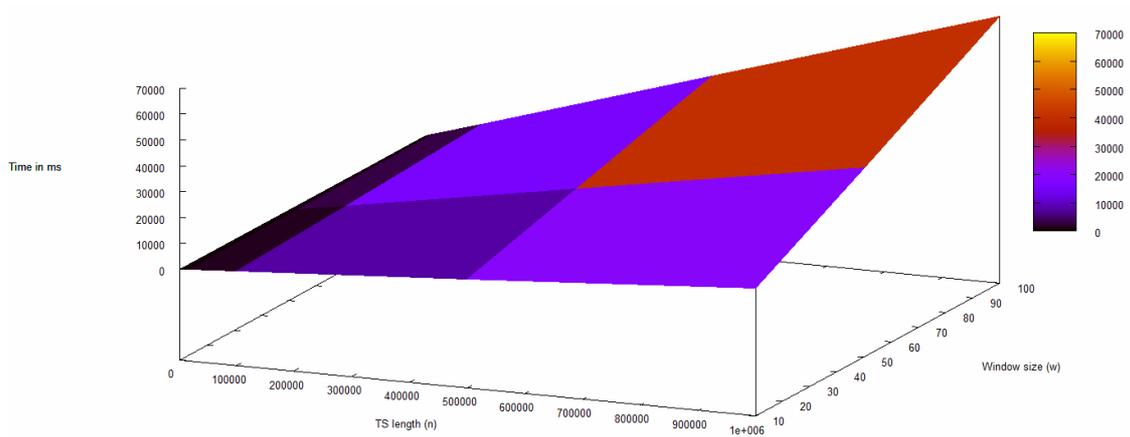

Fig. 3a Java MACD computation time (z) function of TS length (x) and window size (y)

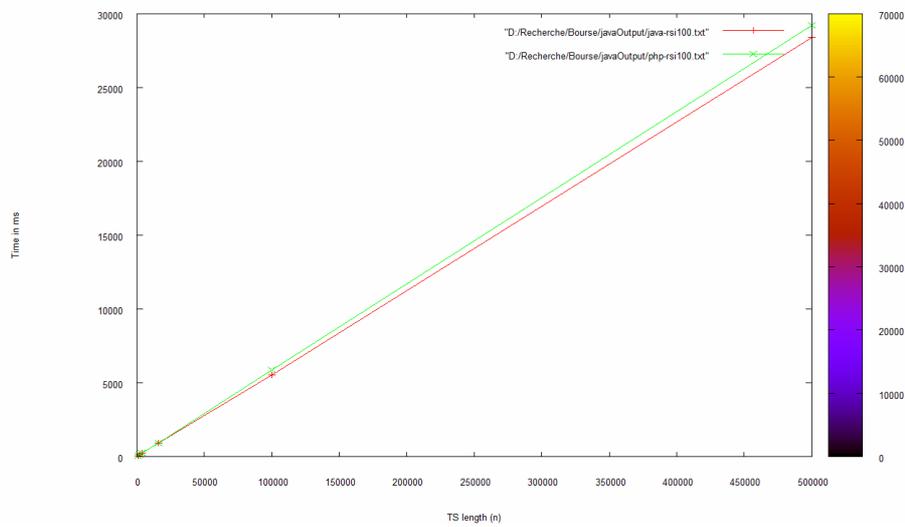

Fig. 3b Java-RSI (+) vs. PHP-RSI (x)

| N | W | PHP | JAVA | QIZX |
|---|---|-----|------|------|
| 1000 | 10 | 25 | 15 | N/A |
| 1000 | 50 | 78 | 40 | N/A |
| 1000 | 100 | 142 | 70 | N/A |
| 2000 | 10 | 51 | 30 | N/A |
| 2000 | 50 | 157 | 80 | N/A |
| 2000 | 100 | 295 | 142 | N/A |
| 4000 | 10 | 103 | 62 | N/A |
| 4000 | 50 | 319 | 164 | N/A |
| 4000 | 100 | 591 | 287 | N/A |
| 16000 | 10 | 412 | 238 | N/A |
| 16000 | 50 | 1276 | 630 | N/A |
| 16000 | 100 | 2364 | 1122 | N/A |
| 100000 | 10 | 2576 | 1402 | N/A |
| 100000 | 50 | 7977 | 3822 | N/A |
| 100000 | 100 | 14781 | 6885 | N/A |
| 500000 | 10 | 12879 | 7203 | N/A |
| 500000 | 50 | 39876 | 19156 | N/A |
| 500000 | 100 | 73901 | 34106 | N/A |

Tab. 3 MACD Computation Time (in ms)

| N | W | PHP |
|---|---|-----|
| 1000 | 10 | 1 |
| 1000 | 50 | 1 |
| 1000 | 100 | 0 |
| 2000 | 10 | 2 |
| 2000 | 50 | 1 |
| 2000 | 100 | 2 |
| 4000 | 10 | 4 |
| 4000 | 50 | 3 |
| 4000 | 100 | 3 |

Tab. 4 PKE computation time (PHP only, in ms)

Results of the four queries are given in Tables 1-4. Due to space limitations we have not included many graphics, but we give an example of cost assessment with Java MACD in Fig. 3a and PHP-RSI vs Java-RSI in Fig. 3b. Other figures are available on demand.

**Results distribution**

Each PHP and Java test was run 1000 times, and results were averaged out. Due to their longer execution time, Qizx queries were only tested 10 times. Note that for readability reasons we have not shown measure deviation in Table 1, but relative error in most measures is about 1% for PHP and Java (i.e. for a measure of 30ms we have values ranging from 27 to 33 ms). There was much more deviation using Qizx, since we can not totally control the application, which is also running with muti-processor support. To try to achieve regular benchmarking, we chose to take the minimum time spent processing the query, over the course of 10 distinct runs. Let us stress that time to load the data from disk into memory is not taken into account.

The point of this section is to show that despite efficient programming and optimizing, a centralized system reaches its limits when computing time series in the contexts mentioned earlier.

**Results analysis**

On the whole, our results show that our PHP and Java implementations are more or less equivalent, and that efficient programming, such as a specific implementation of WAVG in Java leads to up to 1000% speed increase. Qizx on the other hand performs badly, and is much slower, and does not implement many maths functions, making some tests impossible to run. Also note for instance for RSI calculation that the XQuery implementation behaves independently to the $w$ factor, due to the algorithm used. This is one reason why we have chosen to implement specific TS support in our upcoming P2P XQuery prototype [27]. Nevertheless, except for the most basic of operators, computing time series of 500.000 elements takes around a minute, which is too long for an efficient full scale analysis of the CAC40 for instance. Processing such information with XQuery is for the moment unfeasible.

RSI and MACD of TS of a length of 4000 are calculated with our tests in under a second. Behaviour is in general linear with regards to parameters $n$ and $w$. We feel that these results show that there is much to gain by dividing all TS into small (approx. 4000) parts and processing them in a P2P network. We will discuss this improvement in Section 5.

**Main memory limitation**

We received "out of heap space" errors when processing long time series (over 1.000.000 values) with a total heap space allocated of 1GB, therefore centralized processing reaches its limits for long TS.

# 5. P2P Implementation

In this section, we introduce a method for efficiently evaluating operators on time series in a distributed query processing way. In the following, we refer to *peer* when talking about a single, stand-alone computer running our application.

## 5.1 Motivations

The motivations behind the P2P implementation are three fold.

(i) As shown above, due to **main memory limitation**, a peer reaches serious difficulties when computing operations on combinations of lengthy time series. **Reducing time processing** is also a problem for complex statistical operations (e.g., moving exponential average). Distributing data to multiple peers is highly desirable when queries on long time series are submitted to a server. We expect main memory and time processing gains. Streaming of time series could be another approach to save memory on a peer, but it is not the goal of this study which deals with historical data.

(ii) **Sharing base and derived time series** is highly desirable, for example in a trading company. Many users can be interested in the same results, or want to reuse results of operation trees or sub-trees to conduct their own analysis. Therefore, every client may need to download a large number of shared time series, which cost transfer time and main memory. For example, suppose $P$ users want to maintain 1000 different shares history at the minute resolution. Each shared time series is composed of ten years of information, at 360 days per year, 8:30 hours a day, and for every day we need to store five different values every minute (open, max, min, close, volume) using a real (4 bytes). The data size needed to store all this information will be:

$$S = 1000 \times 10 \times 360 \times 60 \times 8.5 \times 5 \times 4 = 34GB \qquad (11)$$

Transferring all data from a server to $P$ clients is very costly. Storing $P$ duplicates wastes storage availability.

(iii) **Processing in parallel operation trees** using data partitioning in the P2P network is a promising way for improving response time. The idea of our P2P implementation is to pool storage and computing capacity of the peers in the network. By distributing the storage of time series on multiple peers, parallel execution of operators is possible. Thus, real-time evaluation of financial strategies can be provided using only ordinary personal computers or laptops available in the network. This might powerfully replace streaming approaches for real time decision. Moreover, the global storage of the community is optimized. Note that such pooling makes complex calculations available to *any* person in the network, even using a simple laptop PC, and no longer restricts use to expensive clusters of PCs, available only to trading companies.

## 5.2 P2P Network Architecture

We have implemented the time series server on top of the  [5] using CHORD and our Java TS implementation. Future work involves testing time series on our P2P XQuery database [27]. From a network communication perspective, DHT-based overlays have gained much popularity in both research projects and real-world P2P applications [16]. DHT networks have proven to be efficient and scalable (most of them guarantee O(logN) scalability) in volatile world area network environments [22]. The infrastructure of our P2P system is provided by the P2PTester. Local data management can be done using a relational or XML DBMS, but in our benchmarking we have loaded all documents into main memory.

 is a real-scale P2P performance measurement platform. It is able to monitor time costs in a P2P environment. We use the  infrastructure to calculate (a) time spent rooting the queries, (b) time spent locally by the time series processor and (c) network time. Note that (b) corresponds to the time measured in Section 4. (c) of course is highly dependant on network traffic. 's architecture is shown in Fig. 4. The module circled in red is the time series specific module. We focus in this experimentation on query evaluation. Note that  can also monitor the cost of document indexing, but this is out of the scope of the paper.

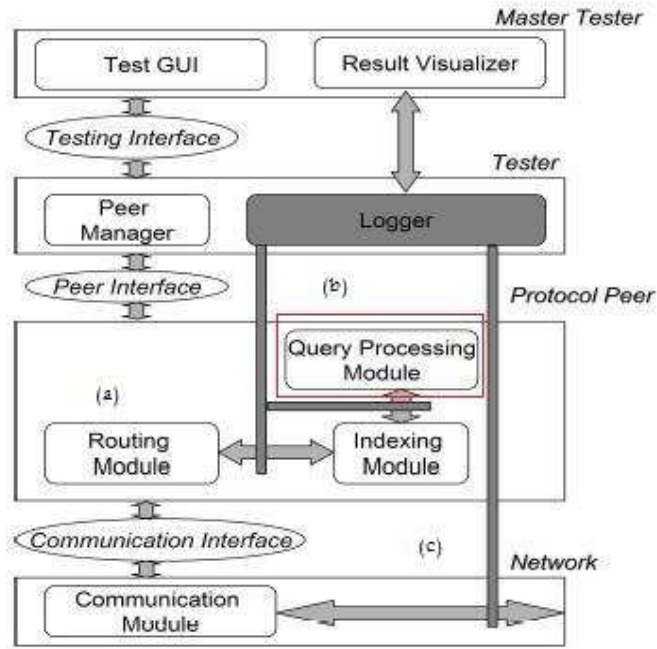

Fig. 4 – architecture

### 5.3 Distributing Time Series to Peers

As time series may be long (e.g., 30 GBs), a peer handling an entire TS might be overloaded, in particular for popular TSs. To avoid this kind of bottleneck, we introduce a method to distribute long TSs into slices on a ring-like addressing space. At loading time, the system distributes time series over the network based on a random hash function. Long time series are split into a sequence of segments. Segments are assigned to peers. Conversely, peers maintain in cache TS segments either imported or calculated. Peers publish the segments they have in cache to other peers by inserting a record in a network DHT. Every segment has the same length (e.g., 1024 entries for stocks). The last segment is in general incomplete and padded with "?" null values. To enable local computation of window-based indicators, we introduce some overlap between segments (e.g., 128 at segment beginning and 128 at segment end for stocks). With such overlap, the local computation of windowing for the core of the series is possible (see Fig. 5).

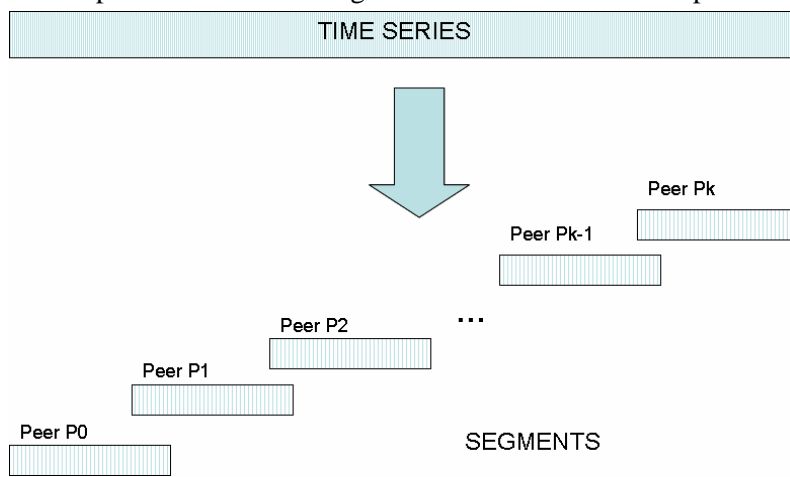

Fig. 5 Distributing time series in segments with overlaps

As explained above, time series reference a calendar. For simplicity, we assume that all time series involved in an application have the same global calendar. This calendar is known by each peer. A calendar is itself a time series giving the *dateTime* for each entry and as such could be distributed in the P2P network. We plan to study this latter. Calendars can be expanded and reduced according to a change of unit (e.g., going from minutes to days). This is done by special operations. This allows changing calendars in query expressions, but not in source time series. Queries are associated to a time interval [<start><end>], where <start> and <end> are indices in the calendar.

A derived time series is described by the attributes name, start, and end. Recall that the name is the functional computation tree of the series. For example, CAC40 could be the name of the base time series representing 20.000 days of the French CAC. MAVG(CAC40, 10) is the derived time series obtained by computing the moving average with a window of 10 days. JOIN(MAVG(CAC40,10), SCALE(MOM(CAC40, 5), 100), SUM) is the join of the previous MAVG and the scaling by 100 of the momentum of the CAC40 with a sliding window of 5, using the SUM mapping function. Thus, the name of a derived series with time interval gives all elements to compute the series but also to retrieve parts of the functional tree computing sub-series. This helps us manage a distributed "semantic" cache of time series as explained below.

A problem is that several functional expressions may compute the same TS, for example SCALE(MOM(CAC40, 5),100) gives the same result that MOM(SCALE (CAC40, 100), 5). This is the classical problem of semantic query rewriting. We define a canonical form of queries to avoid different names for the same derived query. This can be done by introducing rewriting rules useful both for unique functional naming and query optimization; the rules have to include not only basic operations but application operations too. A system of rules shall be convergent towards a unique logical tree in canonical form. However, we leave this for the moment as future work.

Every peer shall retrieve relevant segments of a TS efficiently given a name and a time interval. To reach this goal, a DHT-based index is used. The P2PTester provides a Chord implementation; as required by Chord, the keys are hashed to *m*-bit values in an identifier ring of $2^m$ positions. The P2PTester makes possible to map keys to identifiers in a ring-like addressing space using a specific or a standard hashing function. We select a standard hashing function giving approximately the same probability of hit for each ring node (SHA0). The TS name is selected as a key for the DHT and the publishing peers with associated time intervals are recorded in the DHT entry. Thus, publishing a time series in the network is done by the operation *put(key=<name>, content=(<peerId><start><end>)\*).* Keys are unique, but at each publication of the same key, the list is extended. Other more sophisticated approaches are possible.

To avoid re-computing derived time series, we cache on peers the results of expressions for next uses. We assume each peer has a main memory cache with a replacement policy (e.g., FIFO). A peer loading a base time series segment or producing a derived one keeps the series segment in memory cache if possible. For making it available to other peers, it must publish it on the P2P network. This is simply done by performing a *put* in the Chord network as explained above. Notice that all series computed to materialize a functional tree shall be published if kept in cache. Moreover, when a peer removes a TS segment from its cache, it must remove the corresponding entry from the DHT. Thus, all in all, we introduce a cache-based method for time series query processing in a P2P system described in the next subsection.

## 5.4 Query Processing

At client level, the queries are transformed into functional trees (i.e., logical execution plans) of time series operators. We distinguish atomic sub-queries processing a unique time series (i.e., functional trees composed of SCALE, SEL, PROJ, WIN, MAVG, RSI, etc.) and binary sub-queries (i.e., including PLUS, MINUS, DIVIDE, UNION, INTERSECT, JOIN …). For the time being, N-ary joins are simply translated into a sequence of binary ones; N-ary joins should be supported later for better efficiency. The mediator receiving a query decomposes it into atomic sub-queries followed by binary operations, which gives a primitive canonical form (with a more elaborated canonical form, more hits will be found). Binary operations are processed by the client peer. Atomic sub-queries are searched in the network using the Chord DHT. The *lookup(<name>)* operation of Chord is used for searching available segments published on the peer network. Time intervals have to be assembled to satisfy the query. As base and derived TS can be kept in cache at multiple peers, we need to explore the functional query tree to look for selecting the "best" sub-queries in cache somewhere.

To determine the best set of relevant peers for processing a sub-query, the mediator uses the functional query tree and the time interval asked for (at worst the full calendar, in general a sub-series). To maximize the use of already computed series, the algorithm starts from the functional tree leaves. To each leaf, using the DHT *lookup(key)* function, it determines the relevant peers for computing it. Then, it moves up the tree, trying to find relevant peers for the parent node; if peers caching the functional expression corresponding to the node exist, the children of the leaves are removed. For example, the functional tree calculating the query JOIN(MAVG(CAC40,10), SCALE(RSI(CAC40, 14), 100), SUM) is represented in Fig. 6. A possible selection of peer numbers for the required series is given in brackets, assuming for example that SCALE node is already available as well as the MAVG node. Thus, the two CAC40 nodes and the RSI node can be pruned, as the MAVG and SCALE time series are available on peers [7 and15] and [7] respectively. Notice that, when matching node names, the full name (i.e., the full functional tree with constants if any, encoded as a string), has to be checked. In other words, the $ in Fig. 6 represents the name of the child node, and this recursively. Also note that the query time intervals have to be fully covered. If some parts are missing, they should be recomputed; in other words, the tree has to be inspected segment by segment for all the segments overlapping with the time interval of the query.

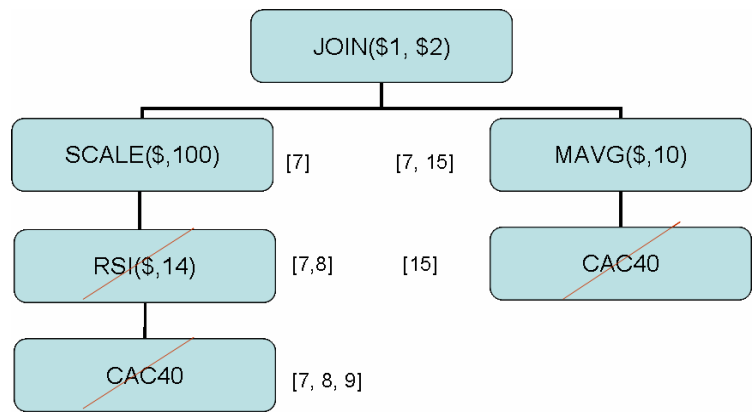

Fig. 6 : Functional tree with annotations (peers) and pruning (lines).

Let us describe the query processing algorithm that has to be executed for each time-segment overlapping with the query time interval. The algorithm performs a depth first traversal of the tree searching for the required time series segment at each node by a DHT lookup(). If the node computation expression (i.e., the TS name) has been published, then the tree node is annotated with the peers handling the relevant TS segments. Fig. 7. gives a sketch of the annotation algorithm. The algorithm also prunes the tree for useless nodes that are replaced by cached nodes. There are several possible choices to select nodes to prune. For the time being, having not developed a cost model, we remove the descendant of any annotated node. When several nodes maintain the searched segment, we select one at random. Here too, the cost model should be helpful.

```
1.  Algorithm Annotated(node)
2.  Input : node a functional tree or a simple leaf.
3.  Output: a functional tree with pruned node.
4.  node.assign ←{}
5.  res ← DHT.lookup(NameComposedPath(node) )
6.  if (res not empty)
7.  then node.assign [peer interval] ← res
8.        if node.left exist    // if unary node
9.      then  delete node.left      // remove all descendants
10.           if node.right exist
11.         then  delete node.right   // same for right side
12.         return
13.
14.    if node.left exist then Annotated(node.left)
15.    if node.right exist then Annotated(node.right)
```

Fig. 7 : Tree annotation and pruning algorithm

## 5.5 Performance Evaluation

In our test setting, we are running on 4 different physical machines, each simulating a certain number of peers $P$. The target time series is split into $P$ different sections, distributed over the peers. We measure $T_{Index}$ the time to find one segment of the TS, $T_R$ the total time to find all segments, $T_P$ the time spent processing the query on each peer, which is found by the experimentation of Section 4.4 (see Tab 3). We also give a calculated estimate of the lower bound of $T_Q$ the time to ship the query, and $T_{NET}$ the time spent for network transfer of results (1GB intranet), which we assume constant given a time series. Note that with a slower connection, time spent in transfer is of course a greater bottleneck. We are running a 500.000 long TS (approx 40 MB) calculating $MACD_{100}$. It is important to note that we consider that the time series is already loaded into the P2P network.

Global cost is equal to: $T_{P2P} = T_R + T_P + T_Q + T_{NET}$ assuming that a peer only has to compute a single segment. Results are shown in Tab. 5. All results are in *ms*.

| P | $T_{INDEX}$ | $T_R$ | $T_P$ | $T_Q$ | $T_{NET}$ | $T_{P2P}$ |
|---|---|---|---|---|---|---|
| 8 | 7,1 | 56,8 | 4473 | <1 | 400 | 4930 |
| 16 | 6,3 | 100,8 | 2176 | <1 | 400 | 2677 |
| 32 | 7,4 | 236,8 | 1106 | <1 | 400 | 1743 |
| 64 | 8,4 | 537,6 | 580 | <1 | 400 | 1518 |
| 128 | 8,9 | 1139,2 | 286 | <1 | 400 | 1825 |
| 256 | 9,7 | 2483,2 | 140 | <1 | 400 | 3023 |

Tab 5. P2P Computation of $MACD_{100}$ of for $n$=500.000

We can see that computation time has a minimum value for $32<P<128$. This value depends on the length of the query, and on the calculation operation, therefore we have not tried to specifically calculate it.

In a centralized client/server environment, total time would simply be $T_{C/S} = T_P + T_{NET}$. If we assume that $T_P=A.n$, $T_R=B.p.\log p$ and $T_{RES}=C.n$ we see that :

$$\frac{T_{P2P}}{T_{CS}} = \frac{B \times p \times \log(p) + A \times \dfrac{n}{p} + C \times n}{A \times n + C \times n} = K_1 \times \frac{p}{n} \times \log(p) + \frac{K_2}{p} + K_3 = \frac{1}{Gain} \qquad (12)$$

We plot in Fig. 8 a simulation of this gain function, with fitted factors, using GnuPlot. As we saw from the results of Tab.5, observe that the maximum gain for $n$=500.000 is obtained for $P \approx 100$. Indeed, this function can be used to predict the number of peers to use for a given TS.

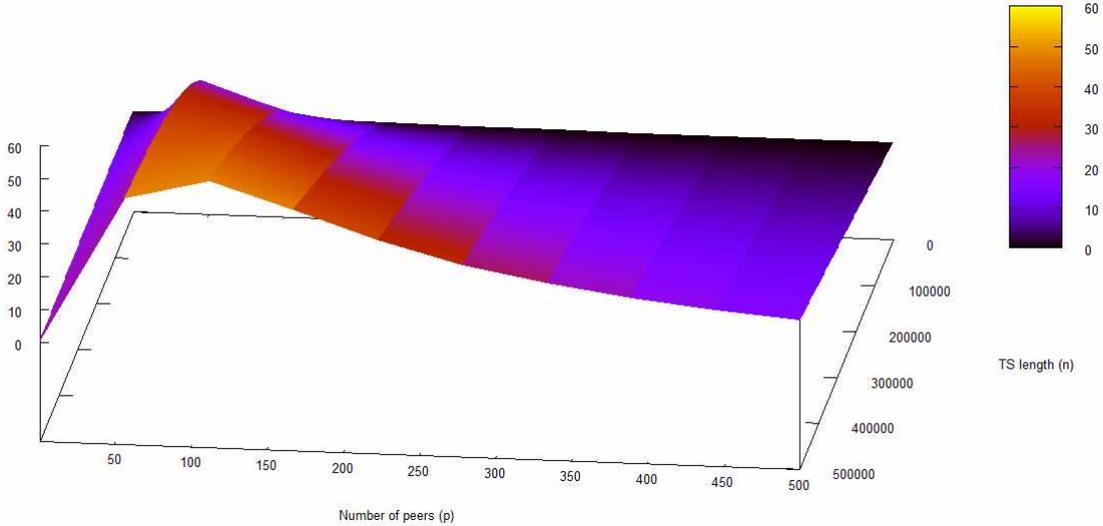

Fig. 8 Relative Gain using P2P, function of number of peers ($P$) and TS length ($n$)

# 6. Related Work

## 6.1 Time Series Functionality

As pointed out in the introduction, a lot of work on time series has focused on efficient representations for distance computation [7]. P2P distance computation could be an interesting new topic that we plan to study. However, as TS are one of the most frequently encountered forms of data, some efforts have been done to efficiently integrate and standardize them in relational DBMSs. The common approach is to introduce the support for sequences as a new ADT. The authors of [21] show the weakness of this approach and propose sorted relations to model sequences, along with a generic algebra and a query optimizer. SQL3 OLAP provides some complex SQL extensions to deal with ordered data and windows. Our approach is more functional, application driven, and extensible. As discussed above, we are geared towards introducing time series in XQuery [28] which already standardizes powerful windowing construct. [17] proposes a complete framework to manipulate and optimize queries on ordered data. The authors develop a new query language for array-tables called AQuery. AQuery could be a good basis for developing a high level query language with a centralized optimizer on top of our TS library. It has to be extended for P2P processing of complex operators, as we propose.

## 6.2 Stream Data Processing

In this article, time series are stored in main memory and not processed on the fly. In contrast to stream data processing [3], we focus on main memory management rather than on real time and continuous queries. Some problems are however common as support of sliding window queries, but the solutions are different. We propose a distributed P2P approach, based on distributed caches. Some papers discuss distributed approaches to stream query processing [6][13]. In particular, [13] proposes parallel execution of stream query functions by partitioning the data stream and processing each partition in parallel, then combining the results. We share with them the idea that to compute efficiently large TS one can slice them. But, the authors of [13] address the problem of continuous queries and do not look at main memory distributed caches in a P2P environment as we do.

## 6.3 P2P DBMS

The vast majority of research on P2P DBMS has concentrated on processing queries on distributed data sources [20][2][19]. Each data source is produced independently by different partners and the role of the P2P networks in a P2P DBMS is to be able to localise relevant collections of data. As data sources are produced independently, many works have addressed the issue of data integration due to the heterogeneity of schemas [26]. In contrast our P2P DBMS works on a horizontal partitioning of time series according to temporality. We do not have the problem of schema integration but our strength is to find every pre-computed node of a functional computation tree on a P2P network. We maximize the use of already computed series that are kept in cache at multiple peers. Compared to other works, such as [19] where the parallelism for query processing is done between relevant collections of data (each collection is located on a given peer), we propose an inner collection (i.e. time series) parallelism. The combination of parallelism inside of a collection and in a set of collections of data may improve the query processing performance.

# 7. Conclusion

In this paper, we have proposed a "stand-alone" extensible model for time series processing. The model is geared towards a main memory implementation of time series. We have compared several implementations, including an integration within an XQuey DBMS. For large time series that do not fit in main memory and/or are time consuming for complex operations, we propose a solution based on slicing time series and sharing slices in a main memory P2P Time

Series Management System (TSMS), and to this end we have developed an initial query optimizer based on functional query plan annotation and distribution.

Two meta-questions may now be considered:

(i) Is the slicing approach efficient enough for amortizing the global overhead of dividing the TS query and assembling the results? We believe that this question can be answered positively. Our benchmark shows that this is possible although the overhead introduced by the network has to be minimized, since it is the greatest bottleneck.

(ii) Is there more into effective P2P time series systems than conventional parallel database technology? We believe that this question can be answered positively. P2P networking brings new problems as DHT indexing and distributed cache management. We provide first solutions to these problems.

To strengthen the affirmative answers to the two questions, we plan to extend our TSMS with more functionality such as support for autoregressive models (ARIMA) and distance computation. These methods require heavy computations, which is a good point for justifying P2P computation and parallelism in general. We plan also to develop a more efficient optimizer for TS queries. We believe that the functional paradigm of our model is well suited for efficient query optimization with efficient cache selection. A cost model to select the best plan is required.

### Acknowledgements

We would like to acknowledge the ROSES project members for many ideas, in particular on the time series model, and the LIVIC Laboratory in Versailles, for the transportation data.